\documentclass[12pt]{article}
\usepackage{epsfig,amssymb,cite,multirow} 
\usepackage{amsmath}

\usepackage{amscd}
\usepackage{amsmath,graphicx}
\usepackage{xypic}
\usepackage[matrix,arrow,curve]{xy}
\usepackage{graphicx}
\usepackage{verbatim}
\usepackage{epsf}
\usepackage{latexsym}
\usepackage{amsmath,amsfonts,amssymb,amsthm}
\usepackage{amsmath,amsthm}

\def\bseq{\begin{subequation}}  % = 1a 1b
\def\eseq{\end{subequation}}
\def\bsea{\begin{subeqnarray}}  % = 1.1a 1.1b
\def\esea{\end{subeqnarray}}
\newcommand{\bbox}{\lower.2ex\hbox{$\Box$}}

\newcommand{\beq}{\begin{equation}}
\newcommand{\eeq}{\end{equation}}
\newcommand{\bea}{\begin{eqnarray}}
\newcommand{\eea}{\end{eqnarray}}
\newcommand{\ena}{\end{eqnarray}}

\newcommand{\e}{\epsilon}

\newcommand{\m}{\mu}

\renewcommand{\)}{\right)}

\newcommand{\be}{\begin{equation}}
\newcommand{\ee}{\end{equation}}

\begin{document}
\setcounter{page}{0}
\begin{titlepage}
\titlepage
\begin{flushright}
LPTENS-10/24\\
UCSD-PTH-10-04\\
\end{flushright}
\begin{center}
\LARGE{\Huge Holographic Optics\\
 and\\
 Negative Refractive Index\\}
\LARGE{\Huge  }
\end{center}
\vskip 1.5cm \centerline{{\bf Antonio Amariti$^{a}$\footnote{\tt amariti@ucsd.edu}, Davide Forcella$^{b}$\footnote{\tt forcella@lpt.ens.fr}, Alberto Mariotti$^{c}$\footnote{\tt alberto.mariotti@vub.ac.be} 
and Giuseppe Policastro$^{b}$\footnote{\tt policast@lpt.ens.fr}
}}
\vskip 1cm
\footnotesize{

\begin{center}
$^a$Department of Physics, University of California\\
San Diego La Jolla, CA 92093-0354, USA
\\
\medskip
$^b$ Laboratoire de Physique Th\'eorique de l'\'Ecole Normale Sup\'erieure \\
and CNRS UMR 8549\\
24 Rue Lhomond, Paris 75005, France
\\
\medskip
$^c$
Theoretische Natuurkunde, Vrije Universiteit Brussel \\
and
The International Solvay Institutes\\ 
Pleinlaan 2, B-1050 Brussels, Belgium\\
\end{center}}

\bigskip

\begin{abstract}
In recent years a very exciting and intense activity has been devoted to the understanding and construction of materials that enjoy exotic optical properties, such as a negative refractive index.  
Motivated by these experimental and theoretical developments, we use the string-inspired idea of holography to study the electromagnetic response of a certain class of media: strongly coupled relativistic systems that admit a dual gravitational description. 
Our results indicate that this type of media generally have a negative refractive index. Moreover we observe that a negative refractive index could be a common feature of relativistic hydrodynamic systems at low frequencies.
\end{abstract}

\vfill
\begin{flushleft}
{\today}\\
\end{flushleft}
\end{titlepage}

\newpage

\tableofcontents

\section*{Introduction}
\addcontentsline{toc}{section}{Introduction}

It is a well-known fact that there exists a close analogy between the free propagation of light
in curved spacetime and the propagation of light inside a material in flat space. 
The geometry bends the light trajectories along null geodesics of the spacetime
metric $g_{\mu \nu}$, while the electromagnetic properties of the medium deviate the trajectory of 
the light according to its electric and magnetic tensors: $\epsilon_{ij}, \mu_{ij}$.
Indeed it is formally possible to rewrite the microscopic Maxwell equations in curved spacetime as 
macroscopic Maxwell equations in a continuos medium in flat spacetime (see for example \cite{Leonhardt:2008}). 

This old story takes a new turn with the recent discovery that one can engineer "real" materials with highly unusual and 
astonishing electromagnetic properties \cite{Smith1,Pendry}. 
They are called metamaterials, since they are 
artificial materials engineered to provide properties which are not readily available in nature. They are typically constructed using periodic arrays of very small and closely spaced elements, 
playing the role of atoms in natural materials. 
One can engineer materials that force the light to follow almost any desired trajectory, allowing the construction 
of cloaking devices, perfect lenses, photonic black holes and so on. We cannot do justice to the literature here, we 
refer the reader to the reviews \cite{Leonhardt:2008, Veselagonew,Ramakrishna,nature,nature2}. 

One of the most interesting properties of some metamaterials is the \emph{negative refractive index}\footnote{Materials with this property are also known as Left-Handed Materials.}.   
The theoretical possibility of this occurrence was first envisaged in 1968 by Veselago \cite{Veselago}, although there are earlier suggestions due to Lamb, Schuster, and Mandelstam, see \cite{Agranovich} and \cite{webpage} for a historical account. 
In such a material the phase of a wave packet propagates in the 
opposite direction to the energy flux, which in turn leads to a number of surprising phenomena: many familiar laws of 
optics are modified, e.g. the Snell's law, the Doppler effect, the Cherenkov effect etc. 
Around 2000 the first samples of negative refractive materials (NRM) were built \cite{Smith1,Pendry,Shelby}  
and since then they have attracted intense interest, due to their multiple technological applications.

It would be important to find new and general classes of physical systems that exhibit negative refraction, and for which it is possible to exactly compute some relevant physical quantities, at least at the linear response level. 
In this paper we observe that the negative refraction seems to be a generic prediction of relativistic hydrodynamics. 
More precisely, we show that if the correlator of the transverse current coupling to the EM field is dominated by a diffusive pole
then there is a low-frequency region where the negative refraction occurs. The presence of the diffusive pole is a consequence 
of the hydrodynamic equations in the case of relativistic field theories at finite temperature 
and chemical potential.

We study in detail 
the refractive index for 
the specific example of a strongly coupled medium (eventually coupled to a dynamical electromagnetic field) 
that admits a gravity dual weakly coupled description.
The computation of the transport coefficients and of the
thermodynamical quantities is performed by
using the string theory-inspired approach of holography.
The holographic analysis goes well beyond the hydrodynamic regime, 
because the response functions are calculated at every frequency.
Our results
show that \emph{certain strongly coupled systems, coupled to a dynamical electromagnetic field, have negative refractive index} in some specified range of frequencies.
This 
confirms the hydrodynamic argument and 
introduces a new connection between gravity and optics, based on the idea of holography, realized in string theory by the AdS/CFT correspondence \cite{Maldacena:1997re}.

The AdS/CFT correspondence is a map between a quantum field theory in flat space, 
and a classical field theory in a completely different spacetime - a curved one, and with a 
larger number of dimensions. Often this is an Anti-de Sitter space or some deformation of it. 
The correspondence is holographic, since one can interpret 
the flat spacetime of the QFT as being the boundary of the curved spacetime. 
In this setup the geometric properties of the higher dimensional curved 
spacetime are translated in properties of matter in flat spacetime (see \cite{Aharony:1999ti} for a review).

In the last decade, this holographic duality has been used with the aim to model several systems 
realized in the laboratories. The first example was the quark-gluon plasma 
produced in heavy-ion collisions \cite{heavy-ion}. 
This was followed by holographic realizations of many phenomena of condensed matter systems: 
superconductivity and superfluidity \cite{superfl},  
Fermi gas at unitarity \cite{fermigas}, the quantum Hall effect \cite{quantumhall}, non-Fermi liquids \cite{nonfermi}, quantum phase transitions \cite{phasetrans}. See \cite{reviewssupercond} for  reviews.

This paper is a first attempt 
towards an optics/geometry holographic duality, 
where string theory can provide a useful
description for some unusual optical properties of electromagnetic media.
 
We believe  that our results could be interesting both for the optics community,
because they offer new theoretical examples and a ``laboratory'' for negative refractive index materials, 
in which the electromagnetic properties of the medium can be computed explicitly; 
and for the string theorists, because they provide a new exciting field of research in 
the AdS/CFT correspondence. 

We should mention at the outset a generic caveat of this string theory setup. 
The systems we are studying are not automatically coupled to a dynamical photon, 
so there is strictly speaking no propagation of light; nevertheless we speak of the refraction index. 
We can do it because, at the first order in the electromagnetic coupling, the optical properties of the medium are completely determined by the linear response to an external electromagnetic field, which can just be taken as a background, non-dynamical field. 
At the phenomenological level we can think of our system as a 
strongly interacting medium weakly coupled to a dynamical photon. The first part of the system is 
studied using the holography, while the coupling to the photon is taken into account perturbatively at the linearized level. 

We should also notice that the systems we study are relativistic, homogeneous, isotropic, strongly-coupled media. We cannot pretend that our results can be 
directly applied to the real metamaterials that in particular are non-relativistic. 
It would be interesting to understand if our results can give 
experimentally testable predictions for systems that occur in nature 
(perhaps in neutron stars or other exotic situations). Moreover it could be interesting to 
study linear the EM response in non-homogeneous, non-isotropic or non-relativistic holographic setups. 

The paper is organized as follows:
in section \ref{linear} we recall the formalism of electromagnetic fields in a medium and the 
linear response theory that allows to derive 
the EM response tensors. In section \ref{main} we 
discuss the conditions under which negative refraction of light occurs, 
and we give our main result: negative refraction is generic 
in a certain class of relativistic systems.
In section \ref{basic} we describe the holographic setup and the 
techniques of the computation. 
In section \ref{numerics} we give more details on the results, then we conclude 
with some additional comments and future 
directions of investigation. 
In the Appendix we discuss some details regarding both the optics and the holographic setup.

\section{Macroscopic EM and linear response theory}
\label{linear}

The analysis of the electrodynamics of continuos media is
traditionally performed by introducing, in addition to the electric and magnetic field, $E$ and $B$, 
the macroscopic fields  $D$ and $H$. They are not independent fields but contain the information about the response of the 
medium to the application of an external field. The general relation, at the linearized level, i.e. for weak external fields, has the form 
\bea{}\label{linEB}
D_i(x,t) = \int \epsilon_{ij} ( t-t') E_j (x,t') dt' \,,\nonumber \\
B_i(x,t) = \int \mu_{ij} ( t-t') H_j (x,t') dt' \,.
\eea
In this approach the properties of the medium are encoded in the (Fourier-transformed) response functions $\epsilon_{ij}(\omega)$, $\mu_{ij}(\omega)$. 
They are often assumed to 
be frequency-dependent only. 
In an isotropic medium, moreover, 
they must reduce to scalar functions: $\epsilon_{ij}(\omega) = \epsilon(\omega) \, \delta_{ij}$,  $\mu_{ij}(\omega) = \mu(\omega) \, \delta_{ij}$. 

If the medium has spatial dispersion 
the response functions also depend on the wave vector $k$.
In this case it is possible  
\cite{Landau}
to reduce the number 
of macroscopic fields to $D$, $E$, $B$, and use the single function 
$\epsilon_{ij}(\omega,k)$ to describe the linear response of the medium:
\begin{equation}\label{epsilonij}
D_i= \epsilon_{ij}(\omega,k) E_j 
\end{equation}
For isotropic media the dielectric tensor $\epsilon_{ij}$ can be decomposed in its transverse and longitudinal part
as
\begin{equation}\label{epsilonLT}
\epsilon_{ij} = \epsilon_T(\omega,k) \left(\delta_{ij}-\frac{k_ik_j}{k^2} \right) + \epsilon_{L}(\omega,k) \frac{k_ik_j}{k^2}
\end{equation}
The dispersion relations are obtained by projecting 
the Maxwell equations on the transverse and longitudinal components. We have
\begin{equation}\label{disprel}
\epsilon_T (\omega,k)=\frac{k^2}{\omega^2}, \quad \quad \epsilon_L(\omega,k) =0
\end{equation} 
In experimental optics the most common description is not in term of $\e_T$ and $\e_L$
but it is in terms of $\epsilon$ and $\mu$. The two approaches can be related in the following way. 
At $k=0$ one has $\epsilon_T = \epsilon_L = \epsilon(\omega)$. 
If we expand the transverse dielectric permittivity as a series in $k$, we can write  \cite{Agranovich}
\begin{equation} \label{agra}
\epsilon_T(\omega,k) = \epsilon(\omega) + \frac{k^2}{\omega^2} \left( 1 - \frac{1}{\mu(\omega)} \right) + {\cal O}(k^4)\,.
\end{equation}
From (\ref{disprel}) and (\ref{agra}), keeping the terms up to $k^2$, we find that 
the transverse dispersion relation can be written as
\begin{equation}
\label{dispersion}
\frac{k^2}{\omega^2}=n^2(\omega) = \epsilon(\omega) \mu(\omegaÊ)
\end{equation}
where $n(\omega)$ represents the refractive index, and this is the same relation usually 
obtained in the $\epsilon - \mu$ approach. 
The two approaches are equivalent when the spatial dispersion is small, and one can neglect the higher-order terms in the 
expansion (\ref{agra}). 

It is important to note that the $\mu(\omega)$ defined via the eq. 
(\ref{agra}) is not the same that appears in (\ref{linEB}). It is instead an effective $\mu$ containing both electric and magnetic effects. 
The magnetic permeability, as commonly defined, can be obtained from the response functions as \cite{Landau} 
\be
1 - \frac{1}{\mu(\omega)} = \omega^2 \hbox{  } \textrm{lim}_{k \to 0} \frac{\epsilon_T (\omega,k)-\epsilon_L (\omega,k)}{k^2} \,.
\ee
However it is the effective $\mu$  that correctly reproduces the dispersion relation and hence the electromagnetic 
propagation and 
dissipation\footnote{Notice that the magnetic permeability $\mu$ can lose its significance at relatively small frequencies \cite{Landau}.}.

The electric permittivity and the magnetic permeability can be obtained from linear response theory \cite{Dressel}. 
In the linear response theory the electromagnetic current $J_i$ is proportional to
the vector potential $A_j$, $J_i= G_{ij} A_j$, where $G_{ij}$ is the retarded correlator 
of the currents in the medium: 
\be{}
G_{ij} (x-x',t-t') = - i \theta(t-t')\,  \langle [ J_i (x,t), J_j (x',t') ] \rangle \,.
\ee
We use for $G$ the same decomposition in transverse and longitudinal part used in eq. (\ref{epsilonLT}), and 
using the macroscopic Maxwell equations we obtain:
\begin{equation}\label{relmax}
\epsilon_T(\omega,k)= 1- \frac{4 \pi}{\omega^2}\hbox{  }q^2 G_T(\omega,k) \,.
\end{equation}
where $q$ is the four-dimensional EM coupling.
Expanding to second order in $k$: $G_T(\omega,k)= G_T^{(0)}(\omega)+ k^2  G_T^{(2)}(\omega)$, we find the electric permittivity and effective magnetic permeability as 
\begin{eqnarray}
\label{fighii}
&&\epsilon(\omega)= 1- \frac{4 \pi}{\omega^2} \hbox{  }q^2 G_T^{(0)}(\omega) \nonumber \\
&&\mu(\omega)=\frac{1}{1+ 4 \pi  \hbox{  }q^2 G_T^{(2)}(\omega)}\simeq  1 - 4 \pi  \hbox{  }q^2 G_T^{(2)}(\omega)
\end{eqnarray}
where in the second line we have expanded for weak EM coupling $q$.
In summary, the response functions, determining the propagation of light in the medium, are given in terms of the retarded correlator of transverse currents\footnote{Longitudinal waves can also propagate if there are solutions of the second equation of (\ref{disprel}), and they could also in principle exhibit negative refraction, but in this paper we will only consider the transverse modes.}.  

\section{Negative refraction}
\label{main}

As we just discussed, the propagation of electromagnetic waves in a medium is usually 
described in terms of the refractive index
 $n$ defined as $n^2 = \epsilon \mu$.
In the absence of dissipation $\epsilon$ and $\mu$ are real. 
In this case, there is wave propagation in the material only if  $\epsilon \mu>0$.
The definition of $n$ does not seem sensitive to the simultaneous change of sign 
of both $\epsilon$ and $\mu$, since
the refractive index is
defined as a quadratic equation.
However, it was understood quite some time ago \cite{Veselago} that changing the sign 
of $\epsilon$ and $\mu$ corresponds to changing the branch in the square root, i.e. 
passing from $n =+ \sqrt{ \epsilon \mu}$, positive refractive index, to $n = - \sqrt{ \epsilon \mu}$, 
negative refractive index. 

The phase velocity of a wave is defined as $v_{ph}= 1/\hbox{Re}(n)$, 
and the change of sign corresponds to the change of 
direction of the phase velocity. 
On the other hand the direction of the Poynting vector, and then of the
energy flow, is not affected by this change of sign.
We are in the 
exotic situation in which the energy flow and the phase velocity are opposite. 
In this case many physical laws change and the material has very special properties such as
inverse Doppler and Cherenkov effect, inverse Snell's law, the medium 
could work as a perfect lens, etc.

However, the situation with constant and negative 
$\epsilon$ and $\mu$ is non physical \cite{Veselagonew}. 
One needs $\epsilon$ and $\mu$ to have a dependence on the frequency. 
In this case of frequency dispersion, 
from general principles we expect $\epsilon$ and $\mu$ 
to acquire an imaginary part, so that the medium is dissipative.
In these media $n$ is itself a complex quantity. Its real part
is the index of refraction while its imaginary part
(usually referred as the extinction coefficient) takes into account the dissipation, and it is always positive in lossy materials.
Differently from the case of real $\epsilon$ and $\mu$, here the 
refraction can be negative also if  Re$(\epsilon)$ and ${Re}(\mu)$
are not simultaneously negative,
because of the presence of the imaginary parts.

For the dissipative case, many different
equivalent conditions for negative refraction have been worked out \cite{McCall,Depine}. 
In 
 \cite{Depine}  (as we review in appendix \ref{wave}) it was shown that the energy flow and the phase velocity are opposite if and only if the index 
\begin{equation}
\label{delfino}
n_{DL}(\omega) = |\epsilon(\omega)| \hbox{Re}(\mu(\omega)) +|\mu(\omega)| \hbox{Re}(\epsilon(\omega)) 
\end{equation}
is negative. 
In the opposite case we have normal wave propagation.
The condition $n_{DL}<0$ 
is equivalent to require that the refractive index is negative. 
In this paper
we will use $n_{DL}$ to check if the medium has 
positive or negative refraction.
\\
\\
It is natural to wonder for which kind of media $n_{DL}<0$.
In the following we study the behavior of $n_{DL}$ in systems that admit 
an hydrodynamic description for 
low frequencies and long wave vector,
and argue  
that there is a simple condition that implies $n_{DL}<0$ in these systems.
We assume the existence of a 
current $J$ such that its transverse part $J_T$ has a diffusive behavior: 
$(\partial_t - \mathcal{D} \nabla^2) J_T = 0 $, where $\mathcal{D}$ is the diffusion coefficient. 
This implies that the retarded correlator of $J_T$ has a diffusive pole, and 
it has the form 
\be\label{hydro}
G_T({\omega,k}) = \frac{i \mathcal{B} \omega}{i \omega - \mathcal{D} k^2}
\ee
with $\mathcal{B}$ a real number, up to contact terms and terms that are subdominant in the hydrodynamic limit (i.e. they are of higher order in $\omega$, $k^2$). 
From generic properties of the Green functions  
we know that: $\mathcal{D} > 0$ (causality implies that the poles can only lie in the lower complex half-plane) and $\mathcal{B}>0$ (Im $G_T < 0$). 
If $J_T$ couples to the EM field with coupling $q$, the electric permittivity and magnetic permeability are
\be\label{hydroem}
\e(\omega) = 1-\frac{4 \mathcal{B} \pi  q^2}{\omega^2}, \quad
\mu(\omega) = 1+\frac{4 i \mathcal{B} \mathcal{D} \pi  q^2}{\omega}
\ee
From these expressions it is easy to show that $n_{DL}<0$ if $\omega^2 < 4 \pi q^2 \mathcal{B}$: 
every system that couples to the EM field with a transverse conserved current 
with retarded correlator 
as in (\ref{hydro}) shows negative refraction at low enough frequencies.

The analysis of the hydrodynamic equations for a relativistic system at finite 
temperature and chemical potential \cite{Kadanoff,ref2hartnoll} 
shows that the transverse current indeed satisfies a diffusion equation; it is also possible to 
relate the constants $\mathcal{B}$ and $\mathcal{D}$ to the values of the transport coefficients and thermodynamical quantities: 
\be \label{coeffi}
\mathcal{B} =  \frac{\rho^2}{\varepsilon+P}, \quad \quad
\mathcal{D} = \frac{\eta}{\varepsilon+P}
\ee
$\varepsilon$ is the energy density, $\rho$ the charge density, $P$ the pressure and $\eta$
the shear viscosity.
Even though it is a simple prediction of hydrodynamics
\footnote{
We are grateful to 
Sean Hartnoll for 
suggesting us 
to investigate the generic structure 
of the correlators dictated by
relativistic hydrodynamics}, 
this observation is new, to our 
knowledge. 
It would be interesting to perform a detailed analysis 
of the EM linear response functions 
for a generic relativistic hydrodynamic system at finite temperature 
and chemical potential, 
and give a general claim about negative
refraction at low frequencies. 
\\
\\
From now on we concentrate on a specific class of systems:
strongly coupled plasmas at finite temperature and chemical potential, 
eventually coupled to a dynamical photon, 
that admit a dual geometric description.
They are very interesting for two main reasons: 
it is possible to explicit compute the transport coefficients and the thermodynamical quantities, and 
it is possible to study the transverse current-current response function for every frequency, going well beyond the hydrodynamic regime. 
Even if they are strongly coupled systems we can perform the
calculations 
by using the holographic approach.
We find a range of frequencies
where the refractive index is negative, for a  
generic class of these materials.
In particular the transverse $G_T({\omega,k})$ 
behaves as (\ref{hydro}) and the constants are given in term of the 
coefficients in (\ref{coeffi}), for low frequencies. 

\begin{figure}
\begin{center}
\includegraphics[width=10cm]{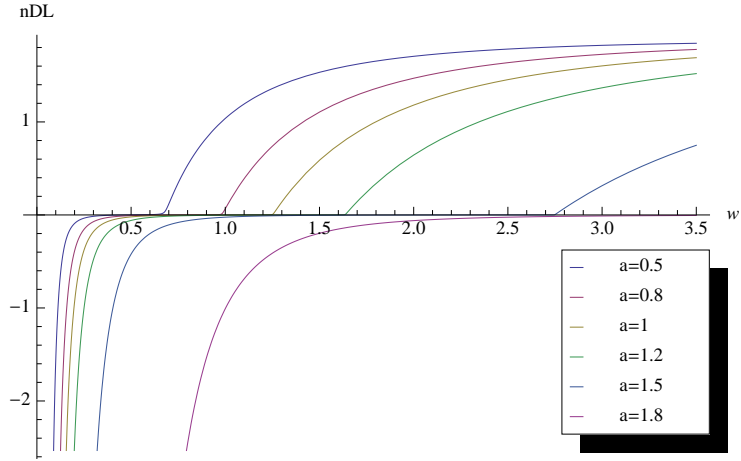}
\caption{$n_{DL}$ for $T=1$ and increasing values of the chemical potential.
We plot $n_{DL}$ 
as a function of the parameter $a$, implicitly defined by the relation 
$\frac{\Sigma}{T} = \sqrt{\frac{6 \pi a}{2-a}}$.
The reason will become clear in the following sections.
}
\label{fighi}
\end{center}
\end{figure}

In Figure \ref{fighi} we show the behavior of the index $n_{DL}$
for fixed temperature and for different values of the 
ratio $\Sigma/T$, where $\Sigma$ is the chemical potential, 
and $T$ the temperature (fixed to $T=1$ for simplicity).%

We observe that,
as expected,
at low frequencies there is negative refraction, whereas for 
high frequencies there is positive refraction.
In the limit of zero chemical potential the index $n_{DL}$
is always positive and there is no negative refraction. 
As the charge is switched on, a region of negative refraction appears. 
This region becomes larger as the charge and/or the temperature increase. 
\\
\\
Before giving some details of the system and the computations, it is important to 
discuss the caveat of the dynamical photon that we already outlined in the introduction.
It is well-known that generically in the holographic correspondence a local 
$U(1)$ symmetry in the bulk corresponds to a global $U(1)$ symmetry in the boundary. 
As a consequence there is no dynamical photon in the dual field theory described 
by the five dimensional geometry. 
As a natural way out we could consider our system as a strongly interacting medium 
weakly coupled to the gauge field associated to the global current we 
are considering. We use the holographic correspondence to describe 
only the medium and we add by hand, in a second time, 
the corrections induced by the presence of a dynamical photon.
We would like to argue that this is a sensible procedure. 
The propagator for a dynamical photon in a given medium is \cite{LandauSP2}:    
\begin{equation}\label{dynphot}
\frac{ 4 \pi}{\omega^2 \epsilon^q_T(\omega,k) - k^2}
\end{equation}
where $q$ is the coupling between the current and the photon: $q J_{\mu} A^{\mu}$, and $\epsilon^q_T(\omega,k)= 1 - \frac{4 \pi}{\omega^2} q^2 G_T^q(\omega,k)$.
$G_T^q(\omega,k)$ is the full retarded current-current correlator. It takes into account the 
interactions among the constituents of the medium and the interactions between the 
medium and the dynamical photon. If the medium and the photon are weakly coupled the 
current-current 
correlator has a natural expansion in power of the coupling constant $q$: $G_T^q(\omega,k)= G_T(\omega,k)+ q^2 G_T^{(2)}(\omega,k) + ...$
In this paper we are able to compute only the first term of this series: $G_T(\omega,k)$.
It is the leading term in perturbative expansion of (\ref{dynphot}) and 
we could be confident that we are considering the leading effects for the propagation 
of a dynamical photon weakly coupled to the strongly coupled medium.
The same philosophy has been applied to the computation of the photo-production rate from the 
quark-gluon plasma in \cite{CaronHuot:2006te}. 

\section{The holographic model }\label{basic}

In this section we introduce our model and we explain in which sense our
optical systems are realized in geometry.

It is in general quite hard to derive $\epsilon(\omega)$, $\mu(\omega)$ for a 
certain given material. In this paper we  take a somehow 
different road and  we compute $\epsilon(\omega)$, $\mu(\omega)$ for a particular class of strongly coupled plasma  coupled to an electromagnetic field. 
The media are characterized by two main parameters: the temperature and the 
chemical potential.

We use the holographic principle to study these systems:
the strongly coupled phase of a quantum field theory living in four dimensional 
Minkowski space time can be described in term of classical gravity in a 
five dimensional Anti de Sitter space (AdS$_5$).
The temperature, and the chemical potential of the field theory are 
encoded in the temperature and charge of a black hole solution in AdS$_5$. 

In this minimal setup our media are described by an energy momentum tensor $T_{\mu\nu}$ and a 
conserved $U(1)$ current $J_{\mu}$.
In the holographic prescription they correspond to a five dimensional metric
$g_{mn}$ and a five dimensional vector $A_{m}$ respectively.
Their dynamics is described by the Maxwell-Einstein action:
\begin{equation} \label{ME}
S = \frac{1}{2 e^2 l^2} \int d^5 x \sqrt{-g} \(R+\!\frac{6}{\, l^2}\) - \frac{1}{4 e^2 } \int d^5 x\sqrt{-g} F_{mn}F^{mn} 
\end{equation}
where $6/l^2$ is the cosmological constant.
The ground state of our systems is described by the particular solution of the equation of motions of (\ref{ME}):
\begin{eqnarray}\label{RN}
&& ds^2= \frac{(2-a)^2 l}{16 \, b^2}\frac{1}{u} \left(\text{d}x^2+\text{d}y^2+\text{d}z^2-f(u) \text{d}t^2 \right)+\frac{l^2}{4 } \frac{\text{d}u^2}{u^2 f(u)} \nonumber \\
&& A_t= - \frac{u}{2 b} \, \sqrt{\frac{3}{2}a}+\Sigma
\end{eqnarray}
where $\Sigma$ is the chemical potential and
\be
f(u) = (1-u)(1+u - a u^2)
\ee 
The solution (\ref{RN}) is a charged black hole in AdS$_5$: the event horizon is at $u=1$, while the
AdS boundary is at $u=0$.   
The chemical potential and the temperature of the system are encoded into the parameter $a$ and $b$
through the relations
\be\label{chemiT}
T = \frac{2-a}{4 \pi b} \, \quad \quad
\Sigma=\frac{1}{2 \, b}\sqrt{\frac{3}{2}a}
\ee 
The propagation of the five dimensional 
photon $A_m$ defines the retarded current-current green function for our (3+1)d
media. 
It is necessary to compute $G_T^{(0)}(\omega)$
and $G_T^{(2)}(\omega)$
 to obtain $\epsilon(\omega)$ and $\mu(\omega)$ defined in the previous section.

The procedure to compute these functions using holography is well-established 
\cite{Son:2002sd}.
To compute the (3+1)d transverse retarded green function $G_T(w,k)$ we need to linearize around the solution
(\ref{RN}) the equations of motions of (\ref{ME}) for the components of the 5d photon $A_m$ transverse 
to the wave vector k. The equations for the photon are coupled to the equations for the metric 
and they decoupled only in the limit of zero charge or zero wave vector ($a \rightarrow 0$ or $k \rightarrow 0$).
We then need to solve these equations imposing the infalling boundary condition 
at the horizon $u=1$. This boundary condition 
correspond to the physical requirement that nothing can escape from the 5d black hole 
at classical level, and it is the retarded prescription for correlators of the (3+1)d medium. From the 
solution for the five dimensional photon we obtain the retarded current-current correlator 
for the media and hence the linear response function $\epsilon(\omega)$ and $\mu(\omega)$.

This system of equation was solved numerically in \cite{Jo:2010sg} and analytically in the hydrodynamic
limit in \cite{ Ge:2008ak,Matsuo:2009yu,Matsuo:2009xn}
\footnote{See also \cite{Hartnoll:2007ip}
for a previous calculation of these Green functions on 2+1 dimensions.}.
The hydrodynamic approximation is a series expansion in $\omega$ and $k$ for the 
equations and solutions.
It is valid only for small values of $\omega$ and $k$, but it offers analytic expressions.
In Appendix \ref{correl} we give the explicit form of the retarded correlator at the lowest order in $\omega$ and $k^2$.
Its leading behavior, necessary for the calculation of $\e(\omega)$ and $\m(\omega)$, is given by 
\be
G^{(0)}(\omega) + k^2 G^{(2)}(\omega) = 
\frac{ l}{4 b (1+a) e^2 } \left( \frac{3 a  }{b}  - i\omega \frac{(2-a)^2    }{2  (1+a)} \right)-\frac{i}{\omega}  
\frac{3 a l}{8  (1+a)^2 e^2 b} k^2 
\ee
Using the equations (\ref{fighii}) we obtain\footnote{We consider a perturbative series expansion
in the (3+1)d EM coupling $q$. The validity of the perturbative expansion would imply a lower bound on the frequencies $\omega$, once all the other physical parameters of the systems are fixed.}:
\begin{eqnarray}
&&\epsilon(\omega)=1+ q^2
\frac{ l \pi  }{b (1+a) e^2 } \left(  \frac{i}{\omega} \frac{(2-a)^2    }{2  (1+a)} - \frac{1}{\omega^2} \frac{3 a  }{b} \right)
 \nonumber\\
&&\mu(\omega)
 \simeq 1 +  q^2 \frac{i}{\omega}  
\frac{3 \pi a l}{2  (1+a)^2 e^2 b}+ \mathcal{O}(q^4)
\end{eqnarray}
\\
The form of $G_T(\omega,k)$, $\epsilon(\omega)$, $\mu(\omega)$, is exactly as in (\ref{hydro}), (\ref{hydroem}), (\ref{coeffi}), except for an higher order term in $\omega$ that does not influence the conclusion of the previous section.
Indeed it is interesting to observe that as soon as we turn on the electric charge of the black hole, i.e. the 
medium has a non vanishing chemical potential, the low frequency behavior of the real part of 
$\epsilon(\omega)$ changes discontinuously to negative value, while the imaginary part of $\mu(\omega)$ 
acquires a positive contribution. 
Even if Re($\mu(\omega)$) never
becomes negative, the dissipation and the fact that Re($\epsilon(\omega)$)$<$0 conspire to give
negative refractive index at low frequencies
\footnote{
The negative quadratic pole in $\epsilon(\omega)$ 
is related at least in part to the translation invariance of the systems we are considering 
\cite{reviewssupercond}. 
Indeed in translationally invariant systems the conductivity $\sigma(\omega)$ has an imaginary pole $i/\omega$, 
and $\epsilon(\omega)=1+ 4\pi i \sigma(\omega)/\omega$. However we have checked that the refractive index is still negative 
even if we add a small amount of impurities that break the translation invariance.}.    
In the next section we will go beyond the hydrodynamic regime and we will compute correlators and response functions for every frequency $\omega$.

\section{Numerics and Plots}\label{numerics}
\begin{figure}
\begin{center}
\includegraphics[width=8cm]{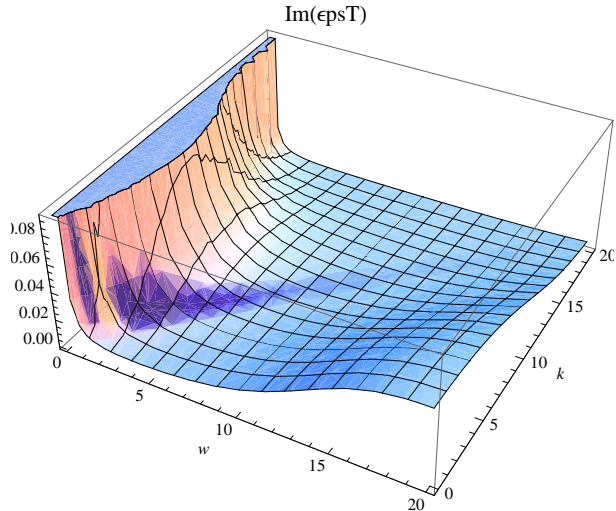}
\caption{Im$\e_T(\omega,k)$ for real $\omega$ and $k$, with $a=1$ and $q=0.05$}
\label{imepsTnok}
\end{center}
\end{figure}
To go beyond the hydrodynamic limit we need to numerically integrate the equations of motion for the 
5d gauge field and the metric.
The functions $\epsilon(\omega)$, $\mu(\omega)$ are obtained from the correlators 
$G_T^{(0)}(\omega)$ and $G_T^{(2)}(\omega)$.
We obtain these Green functions expanding the equations in the wave vector $k$ up to
the order $k^2$ and numerically integrate the resulting equations for every value of $\omega$. As usual the solutions in AdS have IR divergences, that correspond to  UV divergences in the dual field theory, 
that we need to subtract. The holographic renormalization procedure
\cite{Skenderis}
leaves the freedom to add a term to the Green functions of the form $c (\omega^2-k^2)$. 
The constant $c$ represents a renormalization ambiguity that has to be fixed. Our prescription is to  require 
that the electric permittivity approaches  $1$ at large frequencies. This is a natural physical requirement: for infinitely rapid variation of the fields the medium does not have the time to adapt and it behaves like the vacuum.  
    
With these Green functions we can study the optical properties of  the medium that we are describing using gauge/gravity
duality and we can check if it shows negative refraction of light, once coupled to a dynamical photon. 

Here we plot all the results by fixing $T=1$ and by varying the parameters $a$ and 
consequently $b$. We vary $a$ from $a=.5$ to $a=1.8$, so the chemical
potential varies as in (\ref{chemiT}), from $\Sigma\simeq3.5$ to $\Sigma\simeq 52$,
i.e. $\Sigma > T$. Nevertheless at  
higher temperature similar results hold. The regime of negative refraction 
enlarges at higher temperature and reduces at lower ones.

We first have to ensure that the system is in thermodynamical equilibrium.
As we explain in the Appendix \ref{wave},  a system in 
thermodynamical equilibrium must satisfy the requirement 
Im$(\epsilon_T)>0$, and we expect our system to satisfy this
requirements since $\epsilon_T$ is 
given by a Green function. 
We have checked this property by plotting this 
function in Figure \ref{imepsTnok}
for real $\omega$ and $k$.
For this plot we have numerically computed $G_T(\omega,k)$ at all
orders also in $k$.
\begin{figure}
\begin{center}
\begin{tabular}{cc}
\includegraphics[width=8cm]{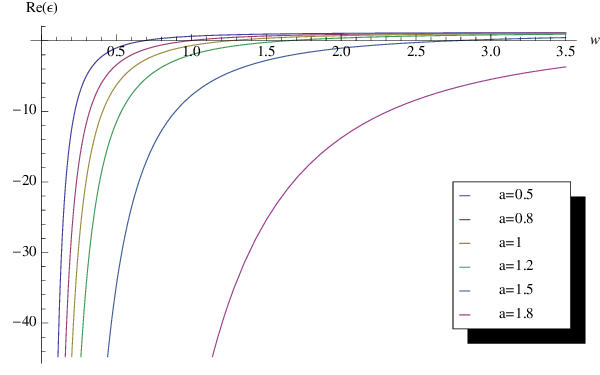}&
\includegraphics[width=8cm]{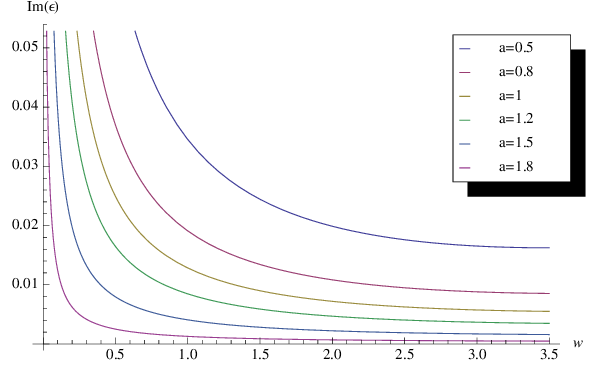}
\end{tabular}
\caption{Re[$\epsilon(\omega)$] and Im[$\epsilon(\omega)$] for different values of $a$ and  $q=0.05$}
\label{fighieps}
\end{center}
\end{figure}
\begin{figure}
\begin{center}
\begin{tabular}{cc}
\includegraphics[width=8cm]{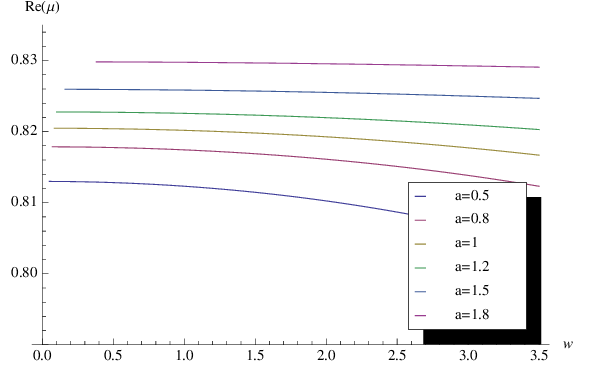}&
\includegraphics[width=8cm]{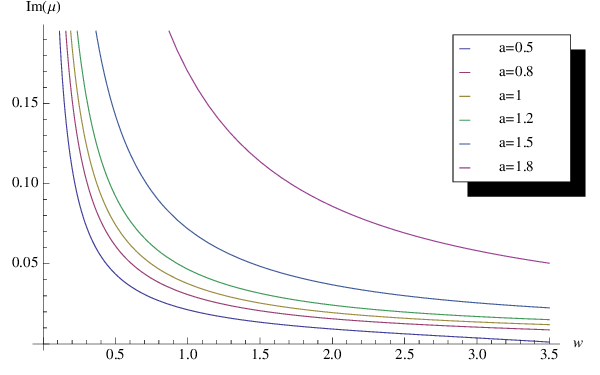}
\end{tabular}
\caption{Re[$\mu(\omega)$] and Im[$\mu(\omega)$]  for different values of $a$ and  $q=0.05$}
\label{fighimu}
\end{center}
\end{figure}

Then we plot the real and imaginary parts of the electric permittivity and of the magnetic
permeability, in Figure  \ref{fighieps} and \ref{fighimu} respectively.
Notice that  Re$(\mu(\omega))$ and Re$(\e(\omega))$ are not simultaneously 
negative. 
As we stressed in 
section \ref{main}, also if Re$(\e)$$\cdot$ Re$(\m)<0$ the energy and phase velocity can be in opposite directions, because of the 
role played  by the imaginary parts, and the material has negative light refraction.

Indeed from Figure \ref{fighi} we see that the index $n_{DL}$ (\ref{delfino}) is negative 
at small frequencies for different values of the parameter $a$.
In the limit  $a \rightarrow 0$ we reduce to the uncharged case, and the index is positive
also at small frequencies.
For small charge $a$, the index becomes immediately negative
at small frequencies, and then it become positive for larger $\omega$. By increasing the value of $a$
the region of negative refraction grows.

\begin{figure}
\begin{center}
\begin{tabular}{cc}
\includegraphics[width=12cm]{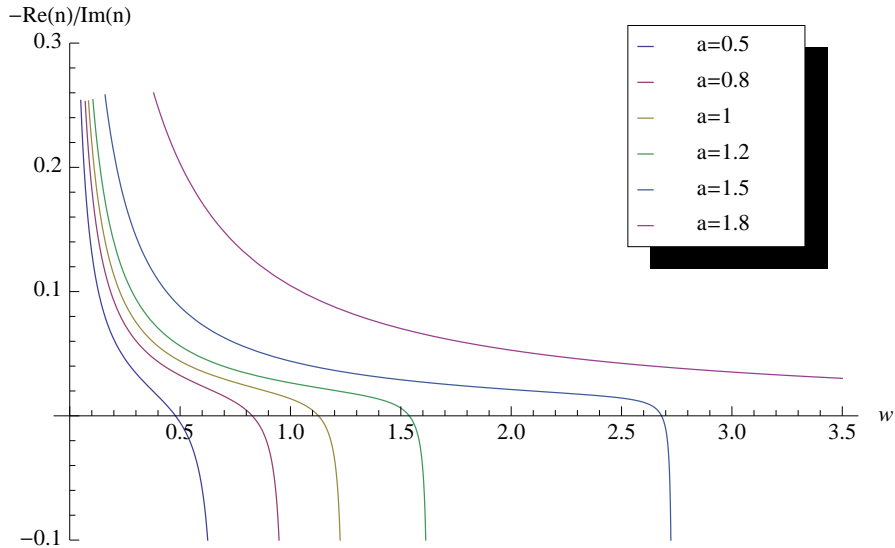}
\end{tabular}
\caption{Ratio -Re[$n(\omega)$]/Im[$n(\omega)$]  for different values of $a$ and  $q=0.05$}
\label{fighin}
\end{center}
\end{figure}

A last useful quantity to plot is the ratio Re$(n)/$Im$(n)$, in Figure \ref{fighin}.
Indeed this quantity  takes into account
the ratio between propagation and dissipation . Usually in isotropic metamaterials it is
common to have strong dissipation in the regime of negative refraction
\cite{Lossy}. This implies
that $\left | \frac{Re(n)}{Im(n)} \right| \ll1 $. Here we see that this is the case too. 
The situation improves as $\omega \rightarrow 0$, where the effects of dissipation get lower. 
At larger frequencies instead the dissipation dominates over the propagation.

As an aside remark, we have found 
that at higher frequencies Im($\mu(\omega)$) becomes negative. 
We stress that it is not a signal of any instability, since  $\epsilon_T(\omega,k)$ is always a well defined response function 
for every value of $\omega$, in particular it always has positive imaginary part.   
We could interpret this phenomenon as the breakdown of the $\epsilon$-$\mu$ approach; it would be desirable  
to have a better understanding of this point.  

\section*{Conclusions and future directions}
\addcontentsline{toc}{section}{Conclusions and future directions}

In the last decade there was a revolution in the practical ability to 
control the electromagnetic fields in various media and the possibility to study
exotic optical phenomena.
In this scenario negative refraction of light has been shown to be 
possible in artificial metamaterials.
One can wonder if there are other physical systems with this
property. In this paper we concentrated on relativistic field theories at
finite temperature and chemical potential. 

We guessed that negative refraction is generic 
in the hydrodynamic limit of these
theories. This behavior is based on the form of the retarded 
propagator of the  transverse current. 
We confirmed this 
observation
with an explicit example.
Indeed we studied the
optical properties of some strongly coupled media 
from the perspective of the gauge/gravity correspondence.
In this case there is a procedure to exactly compute the 
transport coefficients and the response functions 
in terms of the temperature and the 
chemical potential, and we can  obtain 
quantitative informations about the refractive index.

We argued that a charged black hole in AdS$_5$ is a good 
laboratory for the study of
this unusual optical property of three dimensional media.
In this background we have found the range of frequency at which the
index of refraction is negative. 
Even if we studied a specific background, our model is rather generic. 
Indeed the holographic description presented in section \ref{basic} 
is valid for every strongly coupled media with at least a $U(1)$ global symmetry. It is important to stress that we are not studying the optical properties of the complicate five dimensional curved background.  
 This background is only necessary to compute the correlators of the dual strongly coupled 3+1 dimensional field theory, whose  geometry is flat.
Otherwise non-flat geometries lead to other non-conventional optics phenomena (see for example \cite{Hollowood}).

There are many possible future directions of research. 

On a general ground it would be interesting to find if other amazing phenomena
coming from the recent optics developments can be usefully realized in string theory,
or if the ability of string theory to describe some peculiar state of matter could help in optic
devices.

The $U(1)$ gauge symmetry in the five dimensional gravity background is typically understood as
a global symmetry in the dual 3+1 dimensional field theory. We hope that the possibility of negative refraction 
could stimulate the efforts towards a better understanding of the gauging procedure in AdS/CFT 
\cite{Domenech:2010nf}. 
Positive results in this direction could give the possibility to actually compute the first 
quantum electro-dynamical corrections to the refractive index.

As a matter of fact we noted that the dissipation is always 
stronger than the propagation at frequencies corresponding to
negative values of the index of refraction.
It is a common property of many isotropic media \cite{Lossy}, 
and some ad-hoc engineering procedure 
have been used to reduce losses in metamaterials.
It should be interesting to find an analogy of this procedure 
in the system that we studied. 
Maybe one can try to lower the dissipative effects
by studying deformations of the holographic background.

The media we studied in this paper are homogeneous and 
isotropic. To make better contact with real materials it would be interesting to 
break homogeneity and isotropy and to compute current-current correlators in these 
less symmetric backgrounds \cite{Gubser:2009qt,Goldstein:2009cv}. Another  possibility is the breakdown of the relativistic symmetry \cite{Nishida:2007pj,fermigas}.

Negative refraction has been observed also in two dimensional materials \cite{Smith2,Aydin}, and 
it seems natural to extend our analysis of the optical properties from three dimensional 
media to two dimensional ones at weak coupling  and at strong coupling (by using
the AdS$_4$/CFT$_3$  correspondence).

Another natural extension is the study of optical properties of material in the superconducting phase.
Indeed negative refraction has been discussed in superconductors 
in \cite{superconductors}, and maybe the same property can be studied 
in holographic superconductors \cite{superfl} 
(see also \cite{reviewssupercond} and reference therein).

We conclude by noticing 
the 
analogy between negative refraction
in photonics crystals and the motion of the electrons in a bipolar junction in graphene \cite{graphene}.
It can be interesting to better understand this analogy and we hope that our work 
can shed new light on the holographic dual of this effect.

\section*{Acknowledgments}
It is a great pleasure to thank V.~M.~Agranovich, C.~Bachas, F.~Chevy, 
B.~Doucot, Y.~Gartstein, V.~Ginis, S.~Hartnoll, A.~Lakhtakia,
K.~Jo, L.~Mazzanti, Y.~Oz, A.~Salvio, A.~Sen, A.~Starinets, P.~Thassin, 
J.~Troost, I.~Veretennicoff and A.~Zaffaroni 
for nice discussions and clarifications.  
A.A. is supported by UCSD grant DOE-FG03-97ER40546;
A. M.
is a Postdoctoral researcher of FWO-Vlaanderen. A. M. is also supported in part by
the Belgian Federal Science Policy Office through the Interuniversity Attraction Pole IAP
VI/11 and by FWO-Vlaanderen through project G.0428.06;
D. ~F.~ and G.~P.~ are supported by CNRS and ENS Paris. 
A.~A.~, D.~F.~and A.~M.~ thanks the organizers of the Avogadro Meeting and of the 
INFN Vietri conference and the University of Milano-Bicocca where part of this work has been 
done.
G.~P.~ thanks the NORDITA institute where 
part of this work was completed.

\appendix

\section{Wave propagation and dissipation}\label{wave}

In this appendix we derive the conditions for negative refraction and 
the consistency  conditions for the $\epsilon(\omega,k)$ imposed 
by thermodynamics.

All along this paper we use the single function $\epsilon(\omega,k)$ to take into 
account all the interactions between the medium and the electromagnetic field, 
the charges and the currents.
In this setup the Maxwell equations take the following form:
\be
\nabla \cdot D = 0 \qquad, \qquad \nabla \cdot B = 0 \qquad, \qquad \nabla \wedge E = - \partial_t B \qquad, \qquad \nabla \wedge B =  \partial_t D 
\ee
completed by the constitutive relation, written in the dual Fourier space, $D=\epsilon \, E$. The propagation/dissipation of the electromagnetic field in a dispersive/dissipative medium is a bit tricky. 

To determine the propagation of energy in the medium, i.e. the Poynting vector,
we use the approach to consider the field coupled to an external source that 
compensates the energy losses due to the dissipation. In such a way we can think of 
the medium as almost transparent, and the $\epsilon$ is a real function of 
two real coordinates: $\omega,k$. Analyzing the propagation of a wave packet we obtain 
the well-known expression for the Poynting vector in the transparency regime for a dispersive 
medium \cite{Landau}. To take the dissipation of energy into account, we just need to analytically continue the expression 
to complex values for $k$ and $\epsilon$. 
The final expression for the Poynting vector is:
\be{}\label{poip}
{\bf S} = \hbox{Re} \left({\bf E}^* \wedge {\bf B} - \frac{\omega}{2} \frac{\partial \epsilon_{ij}}{\partial {\bf k}} E_i^* E_j \right)  \,.
\ee
The Poynting vector is the energy flow. 
For transverse waves (\ref{poip}) reduces to:
\be
{\bf S} = \hbox{Re} \left(\frac{{\bf k}}{\omega}- \frac{\omega}{2}\frac{\partial \epsilon_T}{\partial {\bf k}}\right) |E_T|^2  \,.
\ee     
If we expand for small values of $k$ as in (\ref{agra}) we obtain the usual expression for the 
Poynting vector in a dissipative medium in the $\epsilon-\mu$ formalism:
\be
{\bf S} = \hbox{Re} \left(\frac{{n}}{\mu} \right) |E_T|^2  \,.
\ee   
where we have used the relation $k /\omega = n$. There is negative refraction when the 
phase velocity is opposed to the flux of energy:
\be\label{egph}
\hbox{Re} \left(n\right)< 0, \qquad \qquad \hbox{Re} \left(\frac{{n}}{\mu} \right)>0
\ee
Working with these relations, 
one can show that (\ref{egph}) is equivalent to $n_{DL}(\omega)<0$. 
This is the condition we use in the main text to check if the refractive index is negative,
which was derived  in 
\cite{Depine}.

Let us now consider the constraints imposed by the thermodynamics on the $\epsilon(\omega,k)$.  
The divergence of $S$ gives the rate of change of the local electromagnetic energy in a medium.
To determine the dissipation we consider a monochromatic wave and we take the divergence
of (\ref{poip}). The first term of (\ref{poip}) gives $\hbox{Re}\left( E^* \partial_t D +
B^* B \right)$, while the divergence of the second term is zero for a plane wave. The 
time average of minus the divergence of the Poynting vector is the energy 
inflow from the external source needed to maintain the plane wave in the medium.
Because in a dissipative medium all the energy of the wave will be eventually transformed 
in heat flow, this quantity is equivalent to the outgoing heat flow; for transverse waves one has:
\be
Q= \hbox{Im} \left( \epsilon_T \right) |E_T|^2
\ee 
Our system, in absence of the external electromagnetic wave perturbation, is 
in thermal equilibrium. Hence the energy flow must be positive: $Q>0$. 
We obtain the consistency condition: $\hbox{Im} \left( \epsilon_T(\omega,k) \right)>0$.
Our epsilon functions are derived from retarded Green functions in a causal, unitary 
field theory for which $\hbox{Im}\left( G(\omega,k) \right) < 0$. For this reason 
the consistency condition for the epsilon is automatically 
satisfied\footnote{The 
inequality holds for $k$ real and $\omega$ lying in the upper complex half-plane; again 
this condition is off-shell from the point of view of  a photon propagating in the medium but we imagine to 
work with an external field that maintains the electromagnetic wave.},
as shown in Figure \ref{imepsTnok}.

\section{Finding the correlators}\label{correl}

In this appendix we describe the calculation of the retarded Green functions that we used in the paper \cite{Jo:2010sg,Ge:2008ak,Matsuo:2009yu,Matsuo:2009xn}.
We need to solve the equations of motion for the fluctuations of the metric and the 
Maxwell field around the solution (\ref{RN}). 
However these
equations are coupled and it is better to define the new variables
\be
\Phi_{\pm} (u)= \frac{1}{u} {h_t^x}'(u) - 3 a B_x(u) +\frac{C_{\pm}}{u}B_x(u) \quad \quad  \quad \small{( \text{with}  \quad
C_{\pm} = (1+a) \pm \sqrt{(1+a)^2 + 3 a b^2 k^2})}
\ee 
$h_t^x(u)$ is a metric fluctuation and $B_x(u) = \frac{A_x(u)}{\Sigma}$,  with $A_x(u)$ a gauge field fluctuation.
The equation of motions for the master variables $\Phi_{\pm}$ are
\begin{eqnarray}
\Phi_{\pm}''(u) +\frac{(u^2 f(u))'}{u^2 f(u)} \Phi_{\pm}'(u) +\frac{b^2}{u f^2(u)}\left(\omega^2 - k^2 f(u) \right) \Phi_{\pm}(u)
-\frac{C_\pm}{f(u)} \Phi_{\pm}(u)= 0 
\end{eqnarray}
It is necessary to solve these equations and find the expression for $B'_x(u)$ to obtain 
the lorentzian correlators.

The analytical solution is found at the lowest order in $\omega$ and $k^2$. 
First we must impose the infalling boundary condition at the horizon $u=1$.
This constrains the solution to be of the form $\Phi_{\pm}(u) = (1-u)^{-\frac{i \omega b}{(2-a)}} \phi_{\pm}$.
The functions $\phi_{\pm}$ are found by solving the equations at the lowest order.
After imposing the  boundary conditions for all the fields at $u = 0$, the derivative of $B_x(u)$ at
the boundary is given by
\be
B_x'(u) =-\frac{ 
\left( \omega k b (h_z^x)^0 + k^2 b (h_t^x)^0 \right)+
3 i a \omega B_0}
{2 i \omega (1+a) -b k^2}
+ i \frac{(2-a)^2 b \omega B_0}{4 (1+a)^2}
+b^2 k^2 B_0 log(u)
\ee
The retarded Green function is obtained from the  boundary action
\be
S \sim - \frac{3 a l }{32 e^2 b^4} \int \frac{d^4 k}{(2 \pi)^4} f(u) B_x (-k,u) B_x'(k,u)
\ee 
by using the prescription of \cite{Son:2002sd} for the minkowskian correlator, and subtracting the UV divergencies:
\be
G_{xx} = \frac{i \omega l}{e^2 } \left( \frac{3 a  }{2 b^2(2 i \omega (1+a) - b k^2)}  - \frac{(2-a)^2    }{8  (1+a)^2 b} \right)
\ee
To obtain a result valid for every value of $\omega$ and $k$, 
we need to implement the above procedure at numerical level.


\begin{thebibliography}{99}

%\cite{Leonhardt:2008}
\bibitem{Leonhardt:2008}
U.~Leonhardt and T.~G.~Philbin,
%``Transformation Optics and the Geometry of Light,''
arXiv:0805.4778v2 [physics.optics]

%\cite{Smith1}
\bibitem{Smith1}
D.~R.~Smith, W.~J.~Padilla, J.~Willie, D.~C.~Vier, S.~C.~Nemat-Nasser. and S.~Schultz, 
%``Composite Medium with Simultaneously Negative Permeability and Permittivity,"
Phys. Rev. Lett. {\bf 84}, 4184-4187 (2000) 

\bibitem{Pendry}
J.~B.~Pendry,  Phys.\ Rev.\ Lett.\, {\bf 85},  3966 (2000)

\bibitem{Veselagonew}
V.~Veselago, L.~Braginsky, V.~Shklover and C.~Hafner, 
%"Negative Refractive Index Materials", 
J. Comput. Theor. Nanoscience {\bf 3}, 1�30 (2006).

%\cite{Ramakrishna}
\bibitem{Ramakrishna}
%``Physics of negative refractive index materials,"
S.~A.~Ramakrishna,  Rep.\ Prog.\ Phys.,\ 2005,  {\bf 68}  449
  
\bibitem{nature}
J.~Valentine, S.~Zhang, T.~Zentgraf, E.~Ulin-Avila, D.~A.~ Genov, G.~ Bartal and X.~Zhang
%"Three-dimensional optical metamaterial with a negative refractive index,"
Nature {\bf 455}, 376-379 ( 2008) 

\bibitem{nature2}
D.~A.~Genov, S.~Zhang and X.~Zhang
%"Mimicking celestial mechanics in metamaterials."
Nature Physics {\bf 5}, 687 - 692 (2009)

\bibitem{Veselago}
V.~G.~Veselago, 
%``The electrodynamics of substances with simultaneously negative values of $\epsilon$ and $\mu$ ,"
Sov. Phys. Usp., {\bf 10}, 509  (1968) 

\bibitem{Agranovich}
  V.~M.~Agranovich, Y.~N.~Gartstein,
  %``Spatial dispersion and negative refraction of light,",
  PHYS-USP {\bf 49} (10), 1029-1044 (2006)
\\
  V.~M.~Agranovich, Y.~R.~ Shen, R.~H.~Baughman and A.~A.~ Zakhidov,
  %``Optical bulk and surface waves with negative refraction,"
  Phys.\ Rev.\  B {\bf 69} (2004) 165112
\\
V.~M.~Agranovich and Y~.N.~Gartstein,
%``Electrodynamics of metamaterials and the Landau�Lifshitz
%approach to the magnetic permeability,''
Metamaterials {\bf 3}  1�9 (2009)

\bibitem{webpage}
http://www.wave-scattering.com/negative.html

\bibitem{Shelby}
R.~A.~Shelby, D.~R.~Smith, S.~Schultz, 
%"Experimental Verification of a Negative Index of Refraction", 
Science {\bf 292}, 77-79 (2001). 

\bibitem{Maldacena:1997re}
  J.~M.~Maldacena,
  %``The large N limit of superconformal field theories and supergravity,''
  Adv.\ Theor.\ Math.\ Phys.\  {\bf 2} (1998) 231
  [Int.\ J.\ Theor.\ Phys.\  {\bf 38} (1999) 1113]
  [arXiv:hep-th/9711200].
  %%CITATION = IJTPB,38,1113;%%
%\cite{Aharony:1999ti}


\bibitem{Aharony:1999ti}
  O.~Aharony, S.~S.~Gubser, J.~M.~Maldacena, H.~Ooguri and Y.~Oz,
  %``Large N field theories, string theory and gravity,''
  Phys.\ Rept.\  {\bf 323} (2000) 183
  [arXiv:hep-th/9905111].
  %%CITATION = PRPLC,323,183;%%
  

\bibitem{heavy-ion}  
  G.~Policastro, D.~T.~Son and A.~O.~Starinets,
  %``From AdS/CFT correspondence to hydrodynamics,''
  JHEP {\bf 0209} (2002) 043
  [arXiv:hep-th/0205052].
  %%CITATION = JHEPA,0209,043;%%
  D.~T.~Son and A.~O.~Starinets,
  %``Viscosity, Black Holes, and Quantum Field Theory,''
  Ann.\ Rev.\ Nucl.\ Part.\ Sci.\  {\bf 57} (2007) 95
  [arXiv:0704.0240 [hep-th]].
  %%CITATION = ARNUA,57,95;%%
  E.~Shuryak,
  %``Physics of Strongly coupled Quark-Gluon Plasma,''
  Prog.\ Part.\ Nucl.\ Phys.\  {\bf 62} (2009) 48
  [arXiv:0807.3033 [hep-ph]].
  
\bibitem{superfl}
  S.~A.~Hartnoll, C.~P.~Herzog and G.~T.~Horowitz,
  %``Building a Holographic Superconductor,''
  Phys.\ Rev.\ Lett.\  {\bf 101} (2008) 031601
  [arXiv:0803.3295 [hep-th]].
  %%CITATION = PRLTA,101,031601;%%
  S.~A.~Hartnoll, C.~P.~Herzog and G.~T.~Horowitz,
  %``Holographic Superconductors,''
  JHEP {\bf 0812} (2008) 015
  [arXiv:0810.1563 [hep-th]].
  %%CITATION = JHEPA,0812,015;%%
  S.~S.~Gubser,
  %``Breaking an Abelian gauge symmetry near a black hole horizon,''
  Phys.\ Rev.\  D {\bf 78} (2008) 065034
  [arXiv:0801.2977 [hep-th]].
  %%CITATION = PHRVA,D78,065034;%%
  C.~P.~Herzog, P.~K.~Kovtun and D.~T.~Son,
  %``Holographic model of superfluidity,''
  Phys.\ Rev.\  D {\bf 79} (2009) 066002
  [arXiv:0809.4870 [hep-th]].
  %%CITATION = PHRVA,D79,066002;%%
  
\bibitem{fermigas}
  D.~T.~Son,
  %``Toward an AdS/cold atoms correspondence: a geometric realization of the
  %Schroedinger symmetry,''
  Phys.\ Rev.\  D {\bf 78} (2008) 046003
  [arXiv:0804.3972 [hep-th]].
  %%CITATION = PHRVA,D78,046003;%%
 K.~Balasubramanian and J.~McGreevy, Phys. Rev. Lett. {\bf 101} , 061601 (2008), arXiv:0804.4053 [hep-th]. 
 
\bibitem{quantumhall}
J.~L.~Davis, P.~Kraus and A.~Shah,
  %``Gravity Dual of a Quantum Hall Plateau Transition,''
  JHEP {\bf 0811}, 020 (2008)
  [arXiv:0809.1876 [hep-th]].
  %%CITATION = JHEPA,0811,020;%%
M.~Fujita, W.~Li, S.~Ryu, and T.~Takayanagi, JHEP {\bf 06}, 
066 (2009), arXiv:0901.0924 [hep-th].
 O.~Bergman, N.~Jokela, G.~Lifschytz, and M.~Lippert,  arXiv:1003.4965 [hep-th].

\bibitem{nonfermi}
  H.~Liu, J.~McGreevy and D.~Vegh,
  %``Non-Fermi liquids from holography,''
  arXiv:0903.2477 [hep-th].
  %%CITATION = ARXIV:0903.2477;%%
 M.~Cubrovic, J.~Zaanen and K.~Schalm,
  %``String Theory, Quantum Phase Transitions and the Emergent Fermi-Liquid,''
  Science {\bf 325} (2009) 439
  [arXiv:0904.1993 [hep-th]].
  %%CITATION = SCIEA,325,439;%%
  T.~Faulkner, H.~Liu, J.~McGreevy and D.~Vegh,
  %``Emergent quantum criticality, Fermi surfaces, and AdS2,''
  arXiv:0907.2694 [hep-th].
  %%CITATION = ARXIV:0907.2694;%%
S.~A.~Hartnoll, J.~Polchinski, E.~Silverstein and D.~Tong,
  %``Towards strange metallic holography,''
  JHEP {\bf 1004} (2010) 120
  [arXiv:0912.1061 [hep-th]].
  %%CITATION = JHEPA,1004,120;%%
  T.~Faulkner, N.~Iqbal, H.~Liu, J.~McGreevy and D.~Vegh,
  %``From black holes to strange metals,''
  arXiv:1003.1728 [hep-th].
  %%CITATION = ARXIV:1003.1728;%%

\bibitem{phasetrans}
 K.~Jensen, A.~Karch and E.~G.~Thompson,
  %``A Holographic Quantum Critical Point at Finite Magnetic Field and Finite
  %Density,''
  JHEP {\bf 1005} (2010) 015
  [arXiv:1002.2447 [hep-th]].
  %%CITATION = JHEPA,1005,015;%%
 K.~Jensen, A.~Karch, D.~T.~Son and E.~G.~Thompson,
  %``Holographic Berezinskii-Kosterlitz-Thouless Transitions,''
  arXiv:1002.3159 [hep-th].
  %%CITATION = ARXIV:1002.3159;%%
 N.~Iqbal, H.~Liu, M.~Mezei and Q.~Si,
  %``Quantum phase transitions in holographic models of magnetism and
  %superconductors,''
  arXiv:1003.0010 [hep-th].
  %%CITATION = ARXIV:1003.0010;%%
  E.~D'Hoker and P.~Kraus,
  %``Holographic Metamagnetism, Quantum Criticality, and Crossover Behavior,''
  JHEP {\bf 1005} (2010) 083
  [arXiv:1003.1302 [hep-th]].
  %%CITATION = JHEPA,1005,083;%%

%\bibitem{reviewssupercond}
%\cite{Hartnoll:2009sz}
\bibitem{reviewssupercond}
  S.~A.~Hartnoll,
  %``Lectures on holographic methods for condensed matter physics,''
  Class.\ Quant.\ Grav.\  {\bf 26} (2009) 224002
  [arXiv:0903.3246 [hep-th]].
  %%CITATION = CQGRD,26,224002;%%
%\cite{Herzog:2009xv}
%\bibitem{Herzog:2009xv}
  C.~P.~Herzog,
  %``Lectures on Holographic Superfluidity and Superconductivity,''
  J.\ Phys.\ A  {\bf 42} (2009) 343001
  [arXiv:0904.1975 [hep-th]].
  %%CITATION = JPAGB,A42,343001;%%
%\cite{Horowitz:2010gk}
%\bibitem{Horowitz:2010gk}
  G.~T.~Horowitz,
  %``Introduction to Holographic Superconductors,''
  arXiv:1002.1722 [hep-th].
  %%CITATION = ARXIV:1002.1722;%%
%\cite{Sachdev:2010ch}
J.~McGreevy,
%Holographic duality with a view toward many-body physics,Ó 
[arXiv:0909.0518 
[hep-th]]. 
  S.~Sachdev,
  %``Condensed matter and AdS/CFT,''
  arXiv:1002.2947 [hep-th].
  %%CITATION = ARXIV:1002.2947;%%

\bibitem{Landau}
L.~D.~Landau, E.~M.~Lifshitz, "Electrodynamics of continous media", Oxford, Pergamon Press, 1984

\bibitem{Dressel}
M.~Dressel and G.~Gruner,
``Electrodynamics of Solids'', Cambridge University Press (2002)

\bibitem{McCall}
 M.~W.~McCall, A.~Lakhtakia and W.~S.~Weiglhofer,
 %The negative index of refraction demystified
 Eur.\ J.\ Phys.,\ {\bf 23} 353

\bibitem{Depine}
R.~A.~Depine and A.~Lakhtakia,
 Microwave and Optical Technology Letters, {\bf 41} 315




\bibitem{Kadanoff} 
L.~P.~Kadanoff and P.~C.~Martin, Annals of
Physics, {\bf 24}, 419 (1963).


\bibitem{ref2hartnoll}
S.~A.~Hartnoll, P.~K.~Kovtun, M.~Muller and S.~Sachdev
%Theory of the Nernst effect near quantum phase transitions in condensed matter and in dyonic black holes
Physical Review B, {\bf 76}, 144502 (2007), [arXiv:0706.3215[cond-mat]]



\bibitem{LandauSP2}
L.~D.~Landau, E.~M.~Lifshitz, "Statistical Physics, Part 2", Oxford, Pergamon Press, 1984 

\bibitem{CaronHuot:2006te}
  S.~Caron-Huot, P.~Kovtun, G.~D.~Moore, A.~Starinets and L.~G.~Yaffe,
  %``Photon and dilepton production in supersymmetric Yang-Mills plasma,''
  JHEP {\bf 0612} (2006) 015
  [arXiv:hep-th/0607237].
  %%CITATION = JHEPA,0612,015;%%


%\cite{Son:2002sd}
\bibitem{Son:2002sd}
  D.~T.~Son and A.~O.~Starinets,
  %``Minkowski-space correlators in AdS/CFT correspondence: Recipe and
  %applications,''
  JHEP {\bf 0209} (2002) 042
  [arXiv:hep-th/0205051].
  %%CITATION = JHEPA,0209,042;%%
  
%\cite{Jo:2010sg}
\bibitem{Jo:2010sg}
  K.~Jo and S.~J.~Sin,
  %``Photo-emission rate of sQGP at finite density,''
  arXiv:1005.0200 [hep-th].
  %%CITATION = ARXIV:1005.0200;%%

%\cite{Ge:2008ak}
\bibitem{Ge:2008ak}
  X.~H.~Ge, Y.~, F.~W.~Shu, S.~J.~Sin and T.~Tsukioka,
  %``Density Dependence of Transport Coefficients from Holographic
  %Hydrodynamics,''
  Prog.\ Theor.\ Phys.\  {\bf 120} (2008) 833
  [arXiv:0806.4460 [hep-th]].
  %%CITATION = PTPKA,120,833;%%

%\cite{Matsuo:2009yu}
\bibitem{Matsuo:2009yu}
  Y.~Matsuo, S.~J.~Sin, S.~Takeuchi, T.~Tsukioka and C.~M.~Yoo,
  %``Sound Modes in Holographic Hydrodynamics for Charged AdS Black Hole,''
  Nucl.\ Phys.\  B {\bf 820} (2009) 593
  [arXiv:0901.0610 [hep-th]].
  %%CITATION = NUPHA,B820,593;%%


%\cite{Matsuo:2009xn}
\bibitem{Matsuo:2009xn}
  Y.~Matsuo, S.~J.~Sin, S.~Takeuchi and T.~Tsukioka,
  %``Magnetic conductivity and Chern-Simons Term in Holographic Hydrodynamics of
  %Charged AdS Black Hole,''
  JHEP {\bf 1004} (2010) 071
  [arXiv:0910.3722 [hep-th]].
  %%CITATION = JHEPA,1004,071;%%



%\cite{Hartnoll:2007ip}
\bibitem{Hartnoll:2007ip}
  S.~A.~Hartnoll and C.~P.~Herzog,
  %``Ohm's Law at strong coupling: S duality and the cyclotron resonance,''
  Phys.\ Rev.\  D {\bf 76} (2007) 106012
  [arXiv:0706.3228 [hep-th]].
  %%CITATION = PHRVA,D76,106012;%%


\bibitem{Skenderis}
K.~Skenderis,
  %``Lecture notes on holographic renormalization,''
  Class.\ Quant.\ Grav.\  {\bf 19} (2002) 5849
  [arXiv:hep-th/0209067].
  %%CITATION = CQGRD,19,5849;%%

\bibitem{Lossy}
%\cite{Stockman}
%\bibitem{Stockman}
M.~I.~Stockman, 
%``Criterion for negative refraction with low optical losses from a fundamental
%principal of causality,''
PRL {\bf 98}, 177404 (2007)

%\cite{Wua}
%\bibitem{Wua}
R.~Wua and D.~Zou,
%``Phase diagram of lossy negative index metamaterials,"
App. \ Phys. Lett. {\bf 93}, 101106 (2008)

%\cite{Zhou1}
%\bibitem{Zhou1}
J.~Zhou, T.~Koschny and C.~M.~Soukoulis,
%``An efficient way to reduce losses of left-handed metamaterials,''
Optical Society of America, Vol. {\bf 16}, No. 15 (2008)

%\cite{Mackay}
%\bibitem{Mackay}
A.~J.~ Mackay,
%``Dispersion requirements of low-loss negative refractive
%index materials and their realisability,''
IET Microw. Antennas Propag., Vol. {\bf 3} , Issue 5, p.808�820  (2009) 

\bibitem{Hollowood}
 T.~J.~Hollowood, G.~M.~Shore and R.~J.~Stanley,
 %``The Refractive Index of Curved Spacetime II: QED, Penrose Limits and Black Holes,"
  JHEP {\bf 0908} (2009) 089  [arXiv:0905.0771 [hep-th]].
  


%\cite{Domenech:2010nf}
\bibitem{Domenech:2010nf}
  O.~Domenech, M.~Montull, A.~Pomarol, A.~Salvio and P.~J.~Silva,
  %``Emergent Gauge Fields in Holographic Superconductors,''
  arXiv:1005.1776 [hep-th].
  %%CITATION = ARXIV:1005.1776;%%
   


%\cite{Gubser:2009qt}
\bibitem{Gubser:2009qt}
  S.~S.~Gubser and F.~D.~Rocha,
  %``Peculiar properties of a charged dilatonic black hole in AdS_5,''
  Phys.\ Rev.\  D {\bf 81} (2010) 046001
  [arXiv:0911.2898 [hep-th]].
  %%CITATION = PHRVA,D81,046001;%%

%\cite{Goldstein:2009cv}
\bibitem{Goldstein:2009cv}
  K.~Goldstein, S.~Kachru, S.~Prakash and S.~P.~Trivedi,
  %``Holography of Charged Dilaton Black Holes,''
  arXiv:0911.3586 [hep-th].
  %%CITATION = ARXIV:0911.3586;%%

%\cite{Nishida:2007pj}
\bibitem{Nishida:2007pj}
  Y.~Nishida and D.~T.~Son,
  %``Nonrelativistic conformal field theories,''
  Phys.\ Rev.\  D {\bf 76} (2007) 086004
  [arXiv:0706.3746 [hep-th]].
  %%CITATION = PHRVA,D76,086004;%%


%\cite{Smith2}
\bibitem{Smith2}
R.~A.~Shelby, D.~R.~Smith, S.~C.~Nemat-Nasser and S.~Schultz, 
%``Microwave transmission through a two-dimensional, isotropic, left-handed metamaterial,"
Appl. Phys. Lett. {\bf 78}, 489 (2001)

%\cite{Aydin}
\bibitem{Aydin}
K.~Aydin, K.~Guven and E.~Ozbay,
%``Two-dimensional Left-handed Metamaterial with a Negative
%Refractive Index,''
Journal of Physics: Conference Series {\bf 36} 6�11 (2006)

\bibitem{superconductors}
%\cite{Pimenov}
%\bibitem{Pimenov}
A.~Pimenov, A.~Loidl, P.~Przyslupski and B.~Dabrowski,
%``Negative Refraction in Ferromagnet-Superconductor Superlattices,''
PRL {\bf 95}, 247009 (2005)

%\cite{Rakhmanov}
%\bibitem{Rakhmanov}
A.~L.~Rakhmanov, V.~A.~Yampol�skii, J.~A.~Fan, F.~Capasso and F.~Nori,
%``Layered superconductors as negative-refractive-index metamaterials,"
Phys. Rev. B {\bf 81}, 075101 (2010)


\bibitem{graphene}
V.~V.~ Cheianov, V.~Fal'ko, and B.~L.~Altshuler
%"The Focusing of Electron Flow and a Veselago Lens in Graphene p-n Junctions,"
Science, Vol. 315. no.{\bf 5816}, pp. 1252 - 1255, (2007)






\end{thebibliography}
\end{document}